\documentclass[12pt]{article}

\usepackage{scicite}

\usepackage{times}
\usepackage{bm}
\usepackage{mathrsfs}
\usepackage{soul}
\usepackage{multirow}
\usepackage{graphicx}
\usepackage{amsmath}
\usepackage{caption}

\newenvironment{sciabstract}{%
\begin{quote} \bf}
{\end{quote}}

\title{Multicolor and 3D holography realized by
\\inverse design of single-celled metasurfaces}

\author
{Sunae So,$^{1,2,\dagger}$ Joohoon Kim,$^{1, \dagger}$ Trevon Badloe,$^{1, \dagger}$ Chihun Lee,$^{1}$ Younghwan Yang,$^{1}$ Hyunjung Kang,$^{1}$ Junsuk Rho$^{1,3,4,5,\ast}$\\
\\
\normalsize{$^{1}$Department of Mechanical Engineering, Pohang University of Science and Technology (POSTECH),}\\
\normalsize{Pohang 37673, Republic of Korea}\\
\normalsize{$^{2}$Graduate School of Artificial Intelligence, Pohang University of Science and Technology (POSTECH),}\\
\normalsize{Pohang 37673, Republic of Korea}\\
\normalsize{$^{3}$Department of Chemical Engineering, Pohang University of Science and Technology (POSTECH),}\\ 
\normalsize{Pohang 37673, Republic of Korea}\\
\normalsize{$^{4}$POSCO-POSTECH-RIST Convergence Research Center for Flat Optics and Metaphotonics,}\\ 
\normalsize{Pohang 37673, Republic of Korea}\\
\normalsize{$^{5}$National Institute of Nanomaterials Technology (NINT),}\\ 
\normalsize{Pohang 37673, Republic of Korea}\\
\\ \normalsize{$^{\dagger}$These authors contributed equally to this work}\\
\\ \normalsize{$^\ast$Corresponding author; E-mail: jsrho@postech.ac.kr}
}

\date{}

\begin{document} 


\baselineskip24pt


\maketitle


\begin{sciabstract}
    Metasurface-generated holography has emerged as a promising route for fully reproducing vivid scenes by manipulating the optical properties of light using ultra-compact devices. However, achieving multiple holographic images using a single metasurface is still difficult due to the capacity limit of a single meta-atom. In this work, we present an inverse design method based on gradient-descent optimization to encode multiple pieces of holographic information into a single metasurface. The proposed method allows the inverse design of single-celled metasurfaces without the need for complex meta-atom design strategies, facilitating high-throughput fabrication using broadband low loss materials. By exploiting the proposed design method, both multiplane RGB color holograms and 3D holograms are designed and experimentally demonstrated. Up to eighteen distinct metasurface-generated holographic images are demonstrated, achieving the state-of-the-art data capacity of a single phase-only metasurface. To the best of our knowledge, we present the first experimental demonstration of metasurface-generated 3D holograms that have completely independent and distinct images in each plane. By demonstrating the high-density holographic information of a single metasurface, the current research findings provide a viable route for practical metasurfaces-generated holography, ultimately stepping towards applications in optical storage, displays, and full-color imaging.
\end{sciabstract}


\section*{Introduction}
    Metasurfaces that exhibit unique optical properties that cannot be found in nature have ushered in a new era of meta-optics with ultra-compact optical devices \cite{yu2011light}. Their constituent components, known as meta-atoms, enable the precise tailoring of electromagnetic waves through arbitrary control of the phase \cite{yu2015high}, amplitude \cite{liu2014broadband}, and polarization \cite{wu2017versatile}. Such properties have been exploited in promising potentials to miniaturize devices in various optical applications such as computer-generated holography \cite{ni2013metasurface,zheng2015metasurface}, imaging \cite{khorasaninejad2016metalenses,chen2018broadband}, and optical communications \cite{zhao2020metasurface,liaskos2018new}. In particular, the versatility of metasurface-generated holography for recreating dynamic images has attracted tremendous interest for various applications in virtual and augmented reality \cite{lee2018metasurface, li2021meta}, optical data storage \cite{wen2015helicity} and security \cite{kim2021pixelated, li2018addressable, kim2022photonic}. For these promising practical applications, it is highly desired to develop multifunctional metasurfaces containing independent holographic information channels for various physical properties of light, including polarization \cite{mueller2017metasurface}, wavelength \cite{wang2016visible}, and orbital angular momentum \cite{ren2020complex}. 
    
\begin{figure}[h]
        \includegraphics[width=140mm]{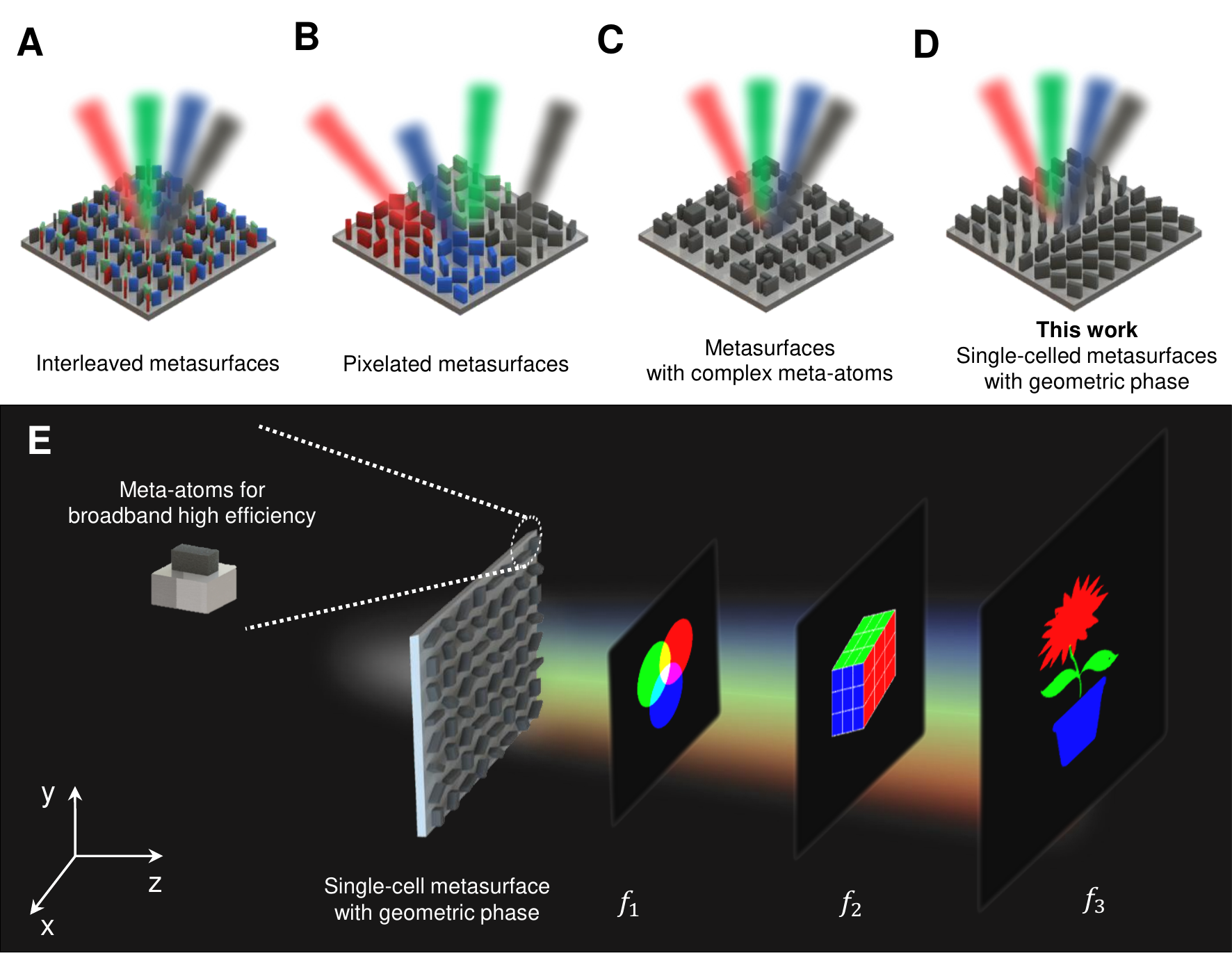}
        \centering
        \caption{Schematic illustration of design strategies for multifunctional metasurfaces based on three previous design strategies of (\textbf{A}) interleaved metasurfaces, (\textbf{B}) pixelated metasrufaces, (\textbf{C}) metasurfaces with complex meta-atoms, and (\textbf{D}) our simple single-celled metasurfaces. (\textbf{A}) The unit-cell of interleaved metasurface consists of several basic sub-components responsible for each functionality. (\textbf{B}) Pixelated metasurfaces are divided into segmented regions, allowing independent light modulation. (\textbf{C}) Other metasurfaces composed of carefully engineered complex meta-atoms. (\textbf{D}) Our single-celled metasurfaces can perform multiple tasks simultaneously by using the geometric phase of simple structures. (\textbf{E}) Schematic illustration of multiplane RGB holograms, one of the examples of multifunctional metasurfaces.} 
        \label{fig:schematics_multifunctional metasurfaces}
    \end{figure}

    The most general approach to design multifunctional metasurfaces is based on physical intuition to circumvent the information capacity limitations of single metasurfaces. These intuition-guided design methods include either using spatial multiplexing schemes \cite{wang2016visible, huang2015aluminum, maguid2017multifunctional} or engineering the meta-atom response to light with different properties \cite{yoon2019wavelength, song2021broadband, xiong2021realizing}. The first approach exploits additional spatial degrees of freedom by using interleaved \cite{wang2016visible, spagele2021multifunctional} or pixelated metasurfaces \cite{huang2015aluminum,song2020ptychography}. Interleaved metasurfaces incorporate multiple optical subcomponents responsible for each function into a minimum operating unit cell (Fig. \ref{fig:schematics_multifunctional metasurfaces}A), while pixelated metasurfaces allow each segmented region to manipulate light independently (Fig. \ref{fig:schematics_multifunctional metasurfaces}B). Although these strategies are the most intuitive ways to achieve multifunctional metasurfaces, they face complications in both the design and fabrication, while facing substantial crosstalk resulting from unwanted coupling between the neighboring structures. Furthermore, the physical size of the unit-cell is vastly increased, which can induce undesired effects such as diffraction. As such, the meta-atoms should be delicately tailored to operate independently to suppress crosstalk in these design strategies, which is accompanied by a significant reduction in efficiency. In addition, the spatial constraints fundamentally restrict the number of functionalities. Recently, multilayer metasurfaces that vertical stack several metasurfaces into a single system \cite{georgi2021optical, zhou2019multifunctional} have been reported to add extra degrees of freedom, but the complexity of the system poses difficulties in fabrication due to the precision needed to align metasurfaces correctly. 
    
    On the other hand, other design strategies for multifunctional metasurfaces have been proposed by carefully designing complex meta-atoms; these approaches often decouple individual functionalities and design delicately tailored meta-atoms to satisfy the desired multiple functionalities independently \cite{dai2020single,overvig2019dielectric,badloe2021electrically}. For example, polarization-multiplexing metasurfaces exploit the polarization-selective responses of asymmetric meta-atoms \cite{arbabi2015dielectric, chen2014high}. The major challenge in designing these multifunctional metasurfaces, however, is finding appropriate meta-atoms to impart desired amplitude and phase independently for different functionalities. An upper limit of six degrees of freedom in a single 2d planar device has been suggested. Based on the Jones matrix of a four meta-atom unit-cell, along with the linear polarization state of the input and output light, three unique near field prints and three far-field holograms have been demonstrated from a single metasurface \cite{bao2021toward}. Recently, modern machine learning schemes have been used to design meta-atoms without decoupling optical functionalities \cite{ma2022pushing,an2021multifunctional}. Still, these methods require a vast library of meta-atoms with complex geometries that result in complicated design and limitations in fabrication (Fig. \ref{fig:schematics_multifunctional metasurfaces}C). In addition, finding meta-atoms may be less practical or even unfeasible as the number of functionalities increases. Most of all, despite the complex design strategies that considered many degrees of freedom, there are limits to experimentally showing multiple holographic images with a single metasurface so far; for example, using an end-to-end design framework, a total of eight holographic images have been achieved experimentally in the near-infrared regime using a single metasurface with complex meta-atoms \cite{ma2022pushing}.
    
    With the ultimate goal to have high-density information of a single metasurface without sacrificing design simplicity and device efficiency, in this work, we present single-celled metasurfaces to achieve multiple holographic images (Fig. \ref{fig:schematics_multifunctional metasurfaces}D). We present an inverse design scheme \cite{so2020deep, molesky2018inverse}  to encode multiple holographic information into a single metasurface through computational optimization. Here, we propose an inverse design method based on gradient-descent optimization method by alleviating optical diffraction calculations and gradient calculation difficulties. This method allows the design and fabrication of metasurfaces comprised of simple meta-atoms by optimizing a single phase profile to provide multiple holographic images. To demonstrate the concept, we design metasurfaces for both multiplane RGB holograms (Fig. \ref{fig:schematics_multifunctional metasurfaces}E), and 3D holography. The phase profiles of the designed metasurfaces are experimentally demonstrated using a high throughput nanoimprint lithography (NIL) fabrication method.

\section*{Results}

\paragraph*{Design principle}

    \begin{figure}[h]
        \includegraphics[width=140mm]{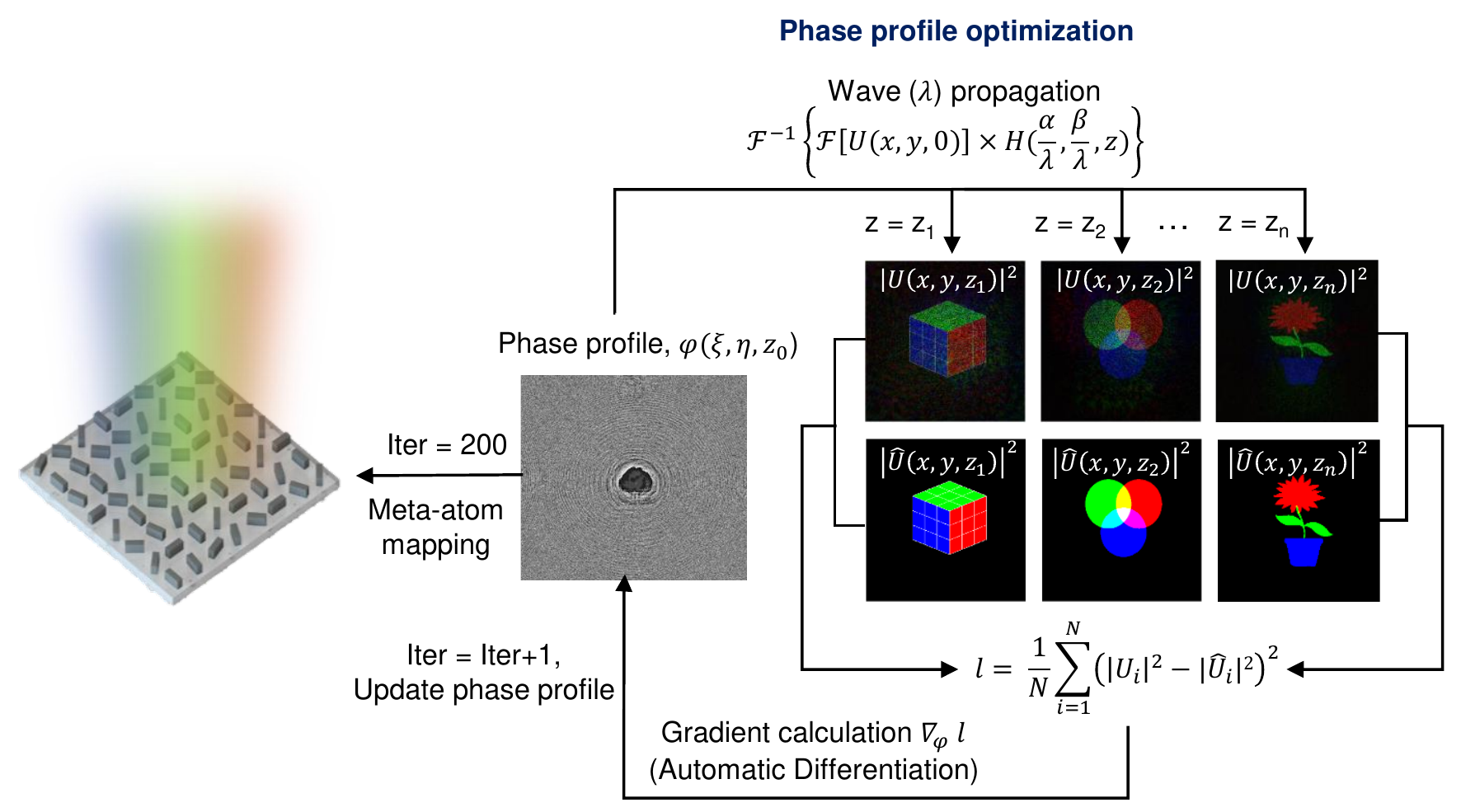}
        \centering
        \caption{Flowchart of the inverse design method for multifunctional metasurface. Starting from a random initial state of phase profile $\varphi$, the phase profile is optimized to minimize the objective function ($l$) that measures the difference between the images reconstructed at the target planes ($z=z_1, ..., z_n$) and the target images. In each iteration, an angular spectrum method is used to compute the forward calculation of the wave propagation, and the optimization is performed by stepping in the opposite direction of the gradient to update the phase profile. For efficient and fast computations, we leveraged automatic differentiation to compute gradients for arbitrary many input variables. After 200 iterations, the optimized phase profile is then mapped to corresponding meta-atoms that impart the required phase profile through rotation of the meta-atoms.}
        \label{fig:schematics}
    \end{figure}
    
     To encode multiple pieces of holographic information into a single phase profile, we present an inverse design method based on gradient-descent optimization (Fig. \ref{fig:schematics}). In the present design method, the goal of inverse design is to optimize the phase profile of the metasurface to have desired holographic images at the target image planes. \textit{i.e.}, the required phase ($\varphi$) at each spatial location of the metasurface is set as an optimization variable and adjusted to minimize the objective function $l$ using a gradient descent optimization based on adaptive moment estimation. The objective function is defined as the mean squared error (MSE) between the target images ($|\hat{U} (x,y,z)|^2$) and the images formed in the image planes ($|U (x,y,z)|^2$), 
    \begin{equation}
         l = \frac{1}{N} \sum_{i=1}^{N} (|(U(x_{i},y_{i},z)|^2 - |\hat{U}(x_{i},y_{i},z)|^2 )^2, 
    \label{eqn:MSE}
    \end{equation}
    where the reconstructed images $|U (x,y,z)|^2$ are calculated using an angular spectrum method (see Supplementary Note 1 for details), 
    \begin{equation}
       U (x, y, z) = \mathscr{F}^{-1}\{ A (\frac{\alpha}{\lambda},\frac{\beta}{\lambda};z)\} = \mathscr{F}^{-1}\{\mathscr{F}\{U(x,y,0)\}\times H(\frac{\alpha}{\lambda},\frac{\beta}{\lambda},z) \}.
    \label{eqn:light_field}
    \end{equation}
    
     For an efficient and fast computation of gradients for arbitrary many input variables, we take advantage of automatic differentiation, a method of calculating partial derivatives for each parameter through the chain rule: within the automatic differentiation framework, the forward problem of light propagation is computed; then, the computation graph is automatically constructed from the forward calculation consisting of a sequence of elementary operations for which all gradients are known. Therefore, by leveraging automatic differentiation, only a single forward simulation is needed to calculate gradients with respect to many input variables. This is in contrast to the traditional brute-force methods that require independent iterative simulations to explore every geometric perturbation with full-wave simulations, prohibiting the overall size of the metasurface to tens of microns \cite{michaels2018leveraging}. Once the gradient ($\nabla_\varphi l$) is known, the optimization is performed by adjusting optimization variables against the gradient. In this work, automatic differentiation and backpropagation are implemented with the aid of the modern machine learning library of PyTorch (See Methods for details).
    
    \begin{figure}[h]
        \includegraphics[width=140mm]{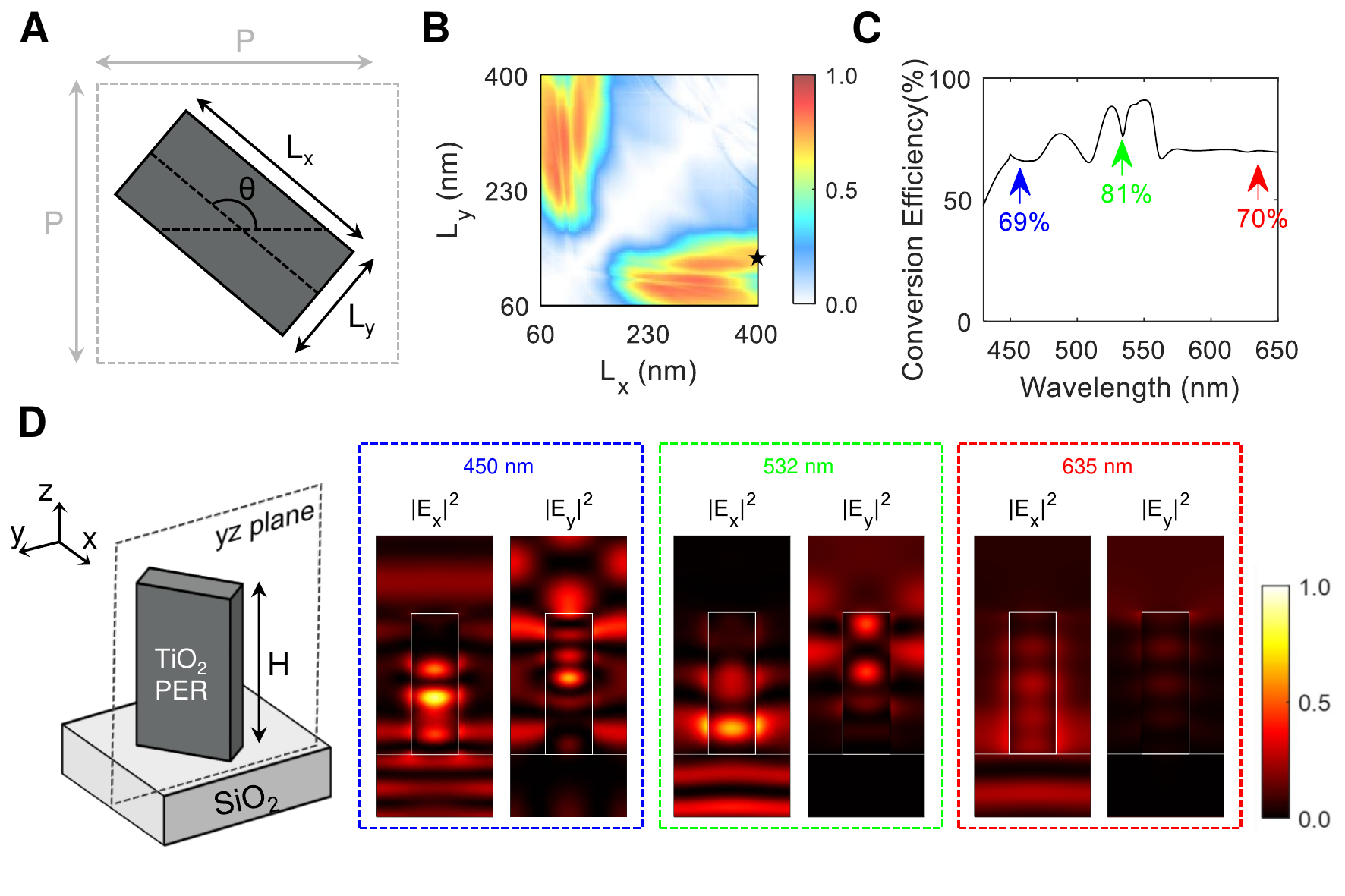}
        \centering
        \caption{Simulation results for unit cell structure design. (\textbf{A}) Top view of the asymmetric nanobar-shaped meta-atoms composed of TiO$_2$ nano PER with broadband low loss properties. The parameters $P, L_x, L_y$, and $\theta$ denote the period, length, width, and the rotation angle of the meta-atom to impart the required geometric phases, respectively. (\textbf{B}) The simulated conversion efficiency with different lengths and widths at $\lambda = 532 \text{nm}$ with fixed period $P=450~\text{nm}$ and height $H=910~\text{nm}$. The meta-atoms with length $L_x=400~\text{nm}$, and width $L_y=120~\text{nm}$ are chosen for the maximum conversion efficiency, as shown in the star-mark. (\textbf{C}) The calculated conversion efficiency of the chosen meta-atoms in the wavelength range from $430~\text{nm}$ to $650~\text{nm}$. For the tree wavelengths of RGB, the conversion efficiency is $70, 81, \text{and} ~ 69 \%$, respectively. (\textbf{D}) The normalized simulated electric field distribution in the yz plane at the three wavelengths, when x-polarized light is incident on the meta-atom rotated by an angle of $\theta = 45 ^{\circ}$.}
        \label{fig:meta-atoms}
    \end{figure}
    
    After the optimization is performed to include the multiple holographic information to the metasurface, the method of physically realizing the phase profile through meta-atoms is required. Since we achieve a single phase profile, we can simply exploit the concept of geometric phase (so called Panacharatnam-Berry phase) using simple asymmetric meta-atoms. This significantly alleviates the need for the design of complex, multi-structured meta-atom. Therefore, asymmetric nanobar-shaped meta-atoms composed of titanium dioxide nano-particle embedded resin (TiO$_2$ nano PER) are used as the building blocks, which provides both ease of fabrication through NIL, and also broadband low loss (Fig. \ref{fig:meta-atoms}, see Supplementary Note 2 for details). To exploit the geometric phase properties, each meta-atom is designed to act as a local half-wave plate that flips the handedness of the input circularly polarized light at the target wavelengths (see Supplementary Note 3 for details). \textit{i.e.}, the meta-atom imparts phase delays of $\pi$ between the two orthogonal electric field components (\textit{i.e.} $|\phi_x - \phi_y| = \pi$). Here, we chose three working wavelengths of $\lambda = 635, 532,$ and $450~\text{nm}$, relating to the three primary colors of red, green, and blue light, respectively. The meta-atoms with height $H=910~\text{nm}$, length $L_x=400~\text{nm}$, width $L_y=120~\text{nm}$, and period $P=450~\text{nm}$ are chosen for the maximum polarization conversion efficiency for three target wavelengths $\lambda = 450, 532,$ and $635~\text{nm}$ (Fig. \ref{fig:meta-atoms}B). The designed meta-atom exhibits a high conversion efficiency over a broad wavelength range of visible light between $400~ \text{and}~ 700~\text{nm}$, with an average efficiency over $70\%$ at three wavelengths (Fig. \ref{fig:meta-atoms}C). Figure \ref{fig:meta-atoms}E shows the electric field distribution in the yz plane when x-polarized light is incident on the meta-atom rotated by an angle of $\theta = 45 ^{\circ}$. It is clear to see that the outgoing wave is polarized along the y-axis, exhibiting the desired half-wave plate-like behavior of the meta-atom at the three target wavelengths. 
    
\paragraph*{Multiplane RGB holograms}
  \begin{figure}[h]
        \includegraphics[width=140mm]{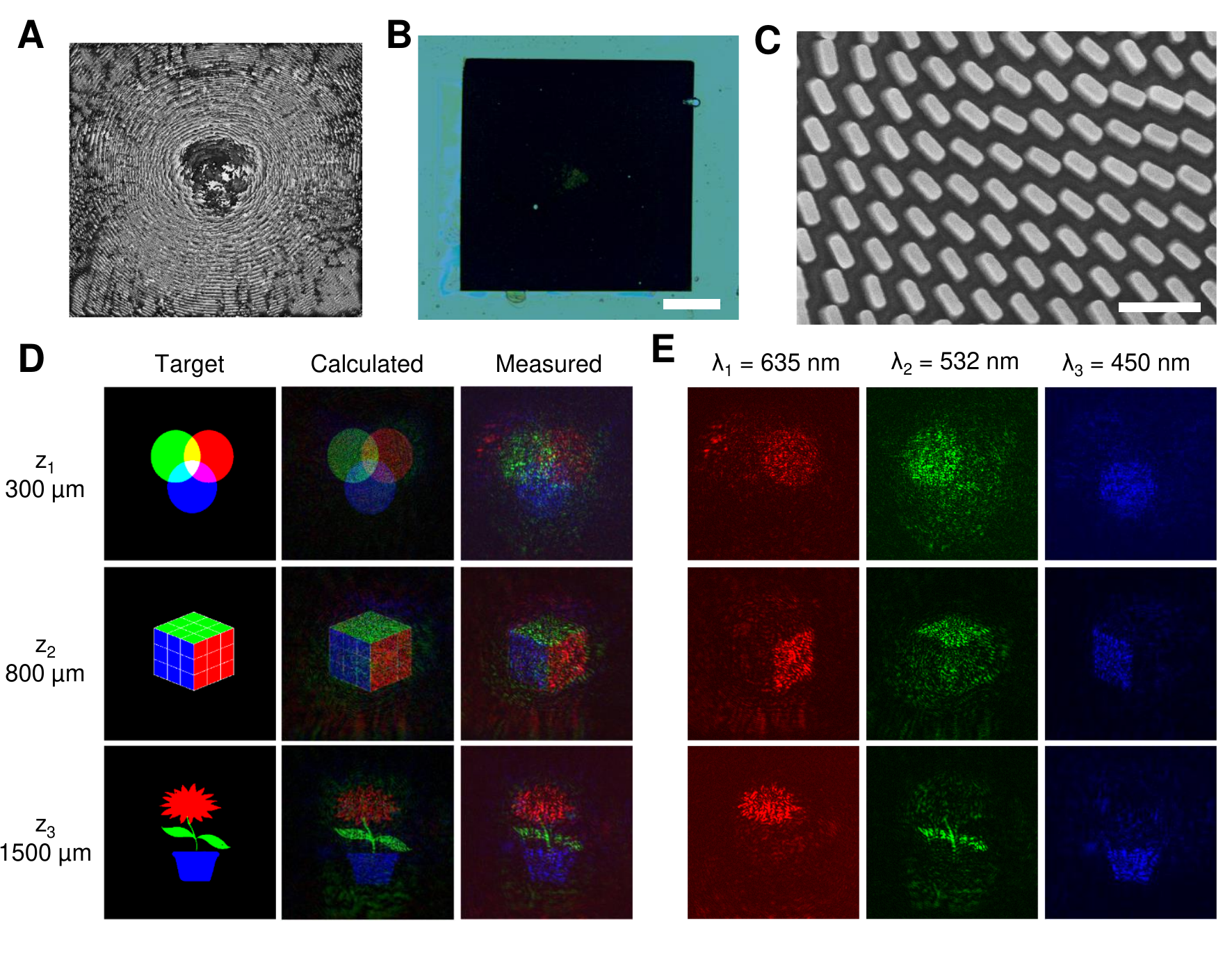}
        \centering
        \caption{Demonstration of multiplane RGB holograms. (\textbf{A}) The phase profile of the optimized metasurface, and (\textbf{B}) and (\textbf{C}) the optical microscopy and scanning electron microscopy images of the fabricated metasurface with the scale bar of $100~\mu\text{m}$ and $1~\mu\text{m}$, respectively. (\textbf{D}) The target, calculated, and experimentally measured holograms are shown at three different focal lengths of $z = 300, 800$, and $1500~\mu\text{m}$. (\textbf{E}) A total of nine individual holographic images are obtained at each of three wavelengths and three focal lengths using a simple single-celled metasurface.}
        \label{fig:holograms}
    \end{figure}

     To verify the design strategy, we conducted the inverse design of a single-celled metasurface to achieve multiplane RGB holograms. Multicolor holograms are one of the most sought-after practical applications of multifunctional metasurfaces but have been difficult to realize due to substantial crosstalk among different wavelengths. Here, we aim to design and fabricate a metasurface consisting of $900 \times 900$ pixels ($405~\mu\text{m} \times 405~\mu\text{m}$) with three focal planes ($300, 800$, and $1500~\mu\text{m}$) corresponding to NA of $0.56, 0.25, 0.13$, respectively. Using the developed inverse design method, the single phase profile for metasurface was optimized to reconstruct target multicolor holograms of the "RGB-color palette," "cube," and "flower" images (Fig. \ref{fig:holograms}A). The designed metasurface was then fabricated using the high-throughput and low-cost fabrication NIL method (Fig. \ref{fig:holograms}B and C, and see Supplementary Note 4 for fabrication details). We emphasize that the designed single-celled metalens can be readily fabricated using such facile, low-cost, repeatable and scalable NIL method because it consists of simple meta-atoms that do not have extremely high-resolution feature sizes. As shown in Figure \ref{fig:holograms}D, the calculated and measured holograms at each focal plane agrees well with the target multicolor holograms (see Supplementary Note 5 for optical setup). We calculate the peak signal-to-noise ratio (PSNR) as a measure of the image quality of the experimentally measured holograms as,
    \begin{equation}
       PSNR = 10\log_{10} \frac{(MAX_I^2)}{MSE},
    \label{eqn:light_field}
    \end{equation}
     where $MAX_I^2$ is the maximum intensity of the target image. The calculated PSNR of the three measured holograms are $12.24$, $12.08$, and $15.56~\text{dB}$, respectively. We observe a relatively high speckle noise in the measured holographic images (Fig. \ref{fig:holograms}D (1)-(9)), especially for the "RGB color palette." This is the typical limitation of phase-only holograms, and the image qualities can be improved by modulating both the phase and amplitude. Nevertheless, despite the information loss due to phase-only holograms, the experimental results show the distinctive multiplane projection capability of the developed metasurface encompassing a total of 9 unique images. Indeed, to the best of our knowledge, this is the highest number of phase-only holograms to have been experimentally demonstrated using a single metasurface, even compared to using complex meta-atom design strategies \cite{ma2022pushing}. 
    
    \begin{table*}[h]
        \footnotesize
        \begin{tabular}{|c|c|c|c|c|c|c|c|}
        \hline
        \multirow{2}{*}{Reference} & \multirow{2}{*}{Materials} & \multicolumn{2}{c|}{Red} & \multicolumn{2}{c|}{Green} & \multicolumn{2}{c|}{Blue} \\\cline{3-8}
         & & $\lambda$ (nm) & Efficiency (\%) & $\lambda$ (nm) & Efficiency (\%)& $\lambda$ (nm) & Efficiency (\%) \\
        \hline
        Ref. \cite{wan2016full} & Al nanoslit & 633 & 1.0  & 532 & 2.8 & 420 & 0.4 \\
        \hline
        Ref. \cite{wen2020multifunctional} & TiO$_2$ &  635 & 4.8  & 532 & 4.7 & 450 & 2.6 \\
        \hline
        Ref. \cite{bao2019full} & c-Si & 633 & 10.3  & 532 & 7.8 & 473 & 6.4 \\
        \hline
        Ref. \cite{wang2016visible} & Si & 633 & 18.0  & 532 & 5.2 & 473 & 3.6 \\
        \hline
        Ref. \cite{hu20193d} & Ag/HSQ/Ag & 633 & 10.8  & 532 & 12.6 & 450 & 22.1 \\
        \hline
        Ref. \cite{jin2018noninterleaved}& Si & 633 & 49.0  & 532 & 30.0 & 488 & 20.0 \\
        \hline
         \textbf{This work} & TiO$_2$ nano PER & 635 & \textbf{61.7}  & 532 & \textbf{75.5} & 450 & \textbf{51.6} \\
        \hline
        \end{tabular}
        \centering
        \captionsetup{justification=centering}
        \caption{Comparison of the measured conversion efficiency for various metasurface-generated multicolor holograms}
        \label{tab:comparison_multicolor holograms}
    \end{table*}
    
    In addition, most design strategies for multicolor holograms to date have resorted to using highly dispersive materials or strong resonances to suppress crosstalk among different wavelengths \cite{wang2016visible}. In contrast, since we can achieve multicolor holograms using a single-phase profile, we can utilize the dispersionless feature of the geometric phase; accordingly, a high- conversion efficiency multicolor hologram is achieved using the broadband low loss of TiO$_2$ nano PER across the entire visible spectrum. The experimentally measured conversion efficiency of the multicolor holograms at R, G, and B are $61.7~\%$, $75.5~\%$, and $51.6~\%$, respectively, where the efficiency is defined as the ratio of the intensity of the light with opposite helicity at the target plane to that of the incident circularly polarized light (see Supplementary Note 7 for details). Table \ref{tab:comparison_multicolor holograms} compares the efficiency of previously reported RGB multicolor holograms. We clearly achieve the highest average efficiency of $63~\%$ at RGB wavelengths. This conversion efficiency is considerably high compared to other works. It should be noted that the conversion efficiency does not directly measure the hologram efficiency, which should be the ratio of the incident power to the desired holograms. Nevertheless, the conversion efficiency still represents the ratio of the energy that is used to generate desired source fields to the incident power. The unprecedented high conversion efficiency is achieved due to the low-loss materials of TiO$_2$ nano PER without significant energy loss. This has hardly been achievable using previously used materials that are highly dispersive or have strong absorptions. The image quality can be further improved by decreasing the number of target holograms or increasing the number of pixels in the metasurface, which comes at a computational cost, however, significantly less than using full-wave simulations (see Supplementary Note 8 for details). 
    
\paragraph*{Experimental demonstration of 3D holograms in multiplane projections}
  \begin{figure}[h]
        \includegraphics[width=140mm]{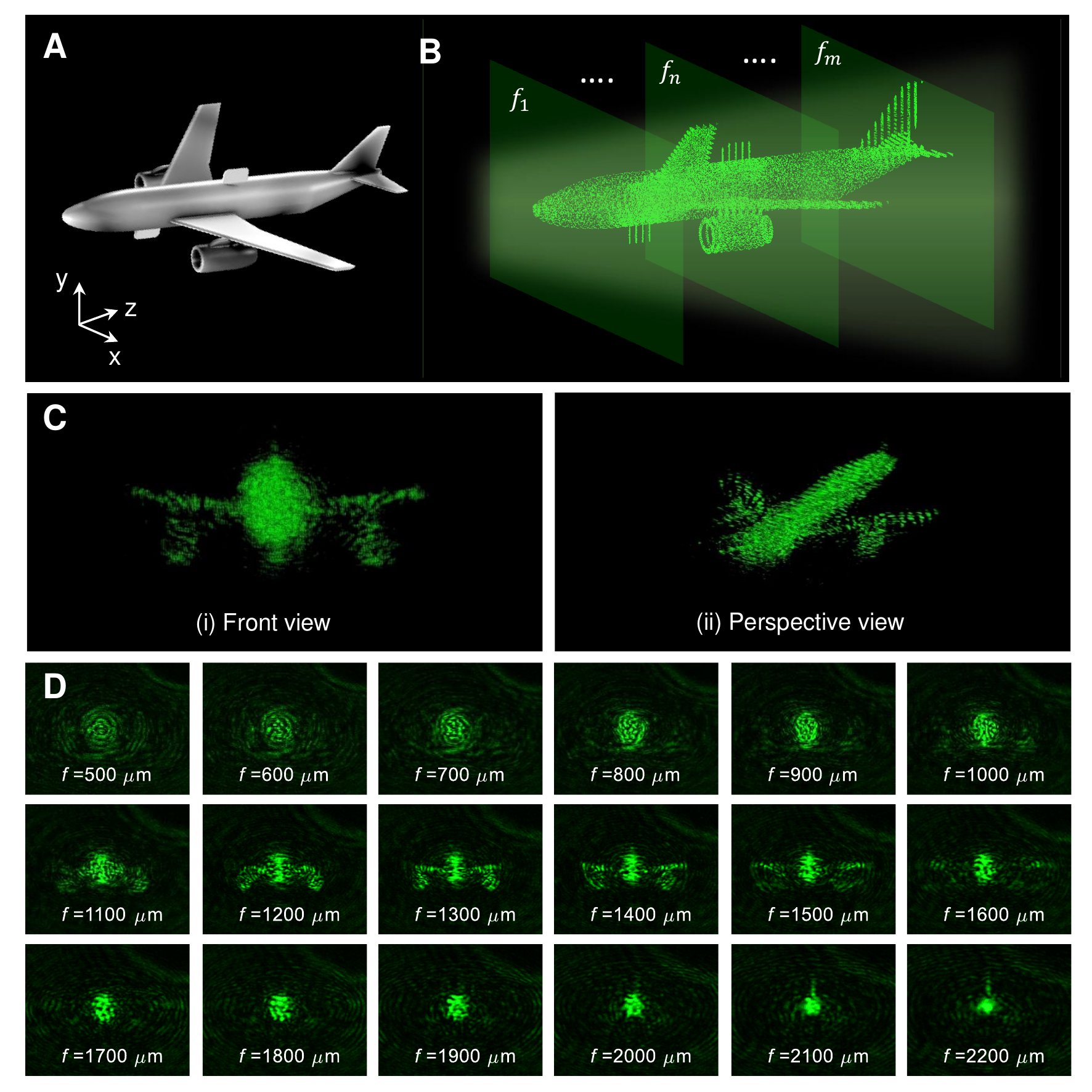}
        \centering
        \caption{Illustration of metasurface-generated 3D hologram. (\textbf{A}) A 3D target object of airplane and (\textbf{B}) its multi-sliced images. (\textbf{C}) The experimentally reconstructed 3D hologram from a single metasurface consisting of $990 \times 990$ pixels. (\textbf{D}) The experimentally measured hologram at each target plane with focal distances linearly spanning from $500~\mu\text{m}$ to $2200~\mu\text{m}$.}
        \label{fig:multiplane_holograms}
    \end{figure}
    
      Furthermore, we present metasurface-generated three-dimensional (3D) holograms. A 3D-like hologram can be achieved by projecting images onto a large number of successive planes. Therefore, for 3D hologram reconstruction, a target object of an airplane (Fig. \ref{fig:multiplane_holograms}A) is sliced into a total of 18 individual images (Fig. \ref{fig:multiplane_holograms}B), and the phase profile for the metasurface was optimized to project the desired image onto each target plane. For the experimental demonstration, a single-celled metasurface consisting of $990 \times 990$ pixels ($455~\mu\text{m} \times 455~\mu\text{m}$) was designed and fabricated to operate at $\lambda = 532~\text{nm}$. Figure \ref{fig:multiplane_holograms}C shows the experimental results of the 3D hologram with a total depth of $1700~\mu\text{m}$. The results clearly demonstrate the successful reconstruction of 3D holographic images with highly suppressed crosstalk. Due to such low crosstalk, holographic images of airplane can be viewed from any observation angle, reconstructing a genuine 3D hologram (See Supplementary Video 1 for experimentally observed 3D holograms). Figure \ref{fig:multiplane_holograms}D shows the on-axis evolution of the total 18 projection planes, which are in good agreement with simulation results (See Supplementary Note 10 for simulation results). To the best of our knowledge, this is the first experimental demonstration of a metasurface-generated 3D hologram having completely independent and distinct images in many successive planes. This is in contrast to other metasurface-generated 3D holography, which present 3D holography by projecting multiple interrelated images onto different planes \cite{li2016multicolor, huang2013three}. By projecting cross-sectional holographic images onto a total 18 image planes, the reconstructed holograms can be observed with a very large viewing angle. This is also remarkable result compared to eleven distinct phase-only holograms experimentally achieved using conventional spatial light modulator at a very low NA of 0.003 to 0.03 \cite{makey2019breaking}. The crosstalk can be further suppressed by increasing the distance between adjacent planes \cite{makey2019breaking}. In addition, by combining the developed inverse design method with other design strategies for wavelength division multiplexing methods, it is expected that we can design multicolor and 3D holograms with an arbitrary number of functionalities.

\section*{Conclusion}
    To conclude, we have proposed an inverse design method to produce simple, single-celled metasurfaces for multicolor and 3D holography. On the basis of the developed gradient-descent optimization method, both multiplane RGB holograms and 3D holograms have been demonstrated by simply exploiting the concept of geometric phase with a single meta-atom. It is worth mentioning that up to eighteen distinct holographic images were experimentally demonstrated without the need for complex design strategies or structures. The simple design also allowed broadband high-efficiency and facilitated facile, low-cost and high-throughput fabrication, paving the way for practical metasurfaces-generated holography. In addition, with efforts to contain high-density data capacity of a single metasurface, the current research findings can be easily combined with other design approaches such as spatial multiplexing, and meta-atom designs, ultimately providing numerous holographic images using a single device. The proposed design method can be generally applied to fields requiring mass information, such as high-density information storage, optical security, and multi-channel images.\\

\section*{Methods}
\noindent\textbf{Gradient descent optimization}\\
    For gradient descent optimization, an adaptive moment estimation (Adam) was implemented as the optimization algorithm,
    \begin{align*}
         m_t = \beta_1 m_{t-1}+(1-\beta_1) \nabla_\varphi l(\varphi_t), \quad & v_t = \beta_2 v_{t-1}+(1-\beta_2) (\nabla_\varphi l(\varphi_t))^2, \quad \varphi_{t+1} = \varphi_t - m_t \frac{\eta}{\sqrt{v_t + \epsilon}},
    \label{eqn:Adam}
    \end{align*}
    where $m$ and $v$ is the first and second moment vector, respectively, and the learning rate of $\eta = 0.05$, and $\beta_1 = 0.9, \beta_2 = 0.999, \epsilon = 1e-08$ were used as hyper-parameters. The Adam optimizer was used to optimize the phase profile over 200 iterations. In this work, the forward calculations of scalar diffraction and phase profile optimizations were performed in Python v3.6 using the PyTorch v1.8.1 framework. \\

\noindent\textbf{Numerical simulation}\\
    To analyze the amplitude and phase response of each meta-atom, numerical simulations using rigorous coupled-wave analysis (RCWA) were performed by using an in-house code. The amplitude and phase retardation of various meta-atoms with height $H=910$ nm, lengths from $L_x=60$ to $400$ nm, width from $L_y=60$ to $400$ nm, and period $P=450$ nm were calculated using the RCWA method. In addition, to quantify the half-wave plate-like behavior of the designed meta-atom, finite-difference time-domain (FDTD) simulations were performed using a commercially available FDTD solver from Lumerical Inc., Ansys. In the FDTD simulations, periodic boundary conditions were used along the x- and y- directions, and perfectly matched layer conditions were used along the z-direction. \\



\bibliography{scibib}

\bibliographystyle{Science}

\section*{Acknowledgments}
    This work was financially supported by the POSCO-POSTECH-RIST Convergence Research Center program funded by POSCO, and the National Research Foundation (NRF) grants (NRF-2019R1A2C3003129, CAMM-2019M3A6B3030637, NRF-2019R1A5A8080290) funded by the Ministry of Science and ICT (MSIT) of the Korean government. S.S. acknowledges the NRF \textit{Sejong} Science fellowship (NRF-2022R1C1C2009430), and the Institute of Information \& Communications Technology Planning \& Evaluation (IITP) grant (No.2019-0-01906, POSTECH Artificial Intelligence Graduate School Program) funded by the MSIT of the Korean government.
 
 \section*{Author contributions}
    S.S., T.B., and J.R. conceived the concept and initiated the project. S.S. developed the method and calculation. S.S., and C.L. performed the numerical calculations. J.K., Y.Y., and H.K. fabricated the metasurfaces and performed the experimental measurements. J.R. guided the entire project. All authors participated in the discussion and confirmed the final manuscript. 
    
\section*{Competing interests}
    The authors declare no competing interests.
    
\section*{Data availability}
    The data that support the findings of this study are available from the corresponding authors upon reasonable request. 
    
\section*{Supplementary materials}
Supplementary Note 1-9\\
Supplementary Video 1\\
Figures S1 to S10\\
Table S1\\
References \textit{(S1-S3)}


\clearpage

\end{document}